\documentclass[aps,twocolumn,showpacs,pre]{revtex4-1}
\usepackage{graphicx,graphics}
\usepackage{mathtools}
\usepackage{amsmath,amssymb,amsfonts}
\usepackage{bbm,bbold}
\usepackage{multirow}
\usepackage{hyperref}
\usepackage{graphicx}
\usepackage{xcolor}

\begin{document}

\title{Dissimilarity between synchronization processes on networks}
\author{Alejandro P. Riascos}
\affiliation{Departamento de F\'isica, Universidad Nacional de Colombia,\\
Bogot\'a, Colombia}

\begin{abstract}
In this study, we present a general framework for comparing two dynamical processes that describe the synchronization of oscillators coupled through networks of the same size. We introduce a measure of dissimilarity defined in terms of a metric on a hypertorus, allowing us to compare the phases of coupled oscillators. In the first part, this formalism is implemented to examine systems of networked identical phase oscillators that evolve with the Kuramoto model. In particular, we analyze the effect of the weight of an edge in the synchronization of two oscillators, the introduction of new sets of edges in interacting cycles, the effect of bias in the couplings, and the addition of a link in a ring. We also compare the synchronization of nonisomorphic graphs with four nodes. Finally, we explore the dissimilarities generated when we contrast the Kuramoto model with its linear approximation for different random initial phases in deterministic and random networks. The approach introduced provides a general tool for comparing synchronization processes on networks, allowing us to understand the dynamics of a complex system as a consequence of the coupling structure and the processes that can occur in it.
\end{abstract}

\maketitle
\section{Introduction}
Synchronization is an emergent collective process in which a  set of coupled agents, under certain conditions, become self-organized evolving to follow the same dynamical pattern \cite{BOCCALETTI2008,VespiBook,PikoBook,strogatzbook,J_Kurths_PhysRep2023}.  This process is one of the most attractive phenomena in complexity science, and some common applications include the synchronization of flashing fireflies \cite{PikoBook,doi:10.1126/sciadv.abg9259}, crowd clapping in a massive event \cite{neda}, and synchronization in arrays of Josephson junctions in condensed matter \cite{josephson}. Synchronization processes occur in extremely diverse systems and play a fundamental role in their functioning \cite{PikoBook,strogatzbook,Balanov2009}. Also, recent research has shown the importance of synchronization in living organisms \cite{BruceWest_2023} and vital functions, e.g., circadian rhythms \cite{WILSON2022,Cascallares_Gleiser_2015}, the functioning of the heart due to the synchronization of pacemaker cells \cite{Mirollo, YANIV20141210} and,  
physiological brain activity \cite{Fell_Axmacher_2011, Wang-X2010, Buzsaki_2006}.
\\[2mm]
In particular, the Kuramoto model is the archetype of collective
systems comprising agents influenced by pairwise nonlinear interactions. Originally introduced to mimic chemical instabilities \cite{Yoshiki_K}, it has become a standard model for the study of the transition to synchrony in agent-based systems and has been applied to diverse complex systems described by networks in contexts such as neuroscience, ecology, and the humanities, among many others \cite{BOCCALETTI2008,Arenas2008,TANG2014184,Rodrigues2016,J_Kurths_PhysRep2023}.
\\[2mm]
Furthermore, the complexity of networks and the varied dynamical processes that may occur in these structures motivate the exploration of measures to quantify the differences between two dynamics. For example, to characterize the effect of a modification in a system or to evaluate how differences in the definition of a mathematical model affect global dynamics. More specifically, in the context of synchronization, the definition of a measure that combines information of the network topology and the dynamics of coupled oscillators may impact diverse applications; for instance, in the study of damage accumulation in a system of coupled identical oscillators \cite{Eraso-Hernandez_2023}, the relation between network structure and synchronizability \cite{Lizier_PNAS_2023},  the effect of stochastic restart \cite{Gupta_Synch_Reset2022}, the detection of the influence of mutual interactions in time-evolving dynamics \cite{Duggento_PRE_2012} or simply to compare the consequences of changing the set of differential equations that define the dynamics \cite{J_Kurths_PhysRep2023}.
\\[2mm]
Here, it is worth noticing that much of the effort in network science has been concentrated on characterizing the topology of networks, with the introduction of a large number of useful measures that allow understanding these structures and their behavior in different applications \cite{VespiBook,Barabasi_2016,Newman_Book2018}. From this information, it is possible to detect similarities or differences between networks by direct comparison of the quantities that describe them. In contrast, there are very few studies aimed at understanding measures that allow the comparison of a network including a specific process.  Recent efforts to compare dynamical processes on networks include distances between networks that usually fall into one of two general categories defining structural and spectral distances, often considered mutually exclusive \cite{Donnat_2018}. The first captures variations in the local structure, as examples of this metric are the Hamming \cite{Hamming1950} and Jaccard distances \cite{Jaccard1901,LEVANDOWSKY1971} characterizing the number of edge deletions and insertions necessary to transform one network into another, more recent metrics used an information theory approach of the entire network structure \cite{Bagrow2019}. In contrast, the spectral approach assesses the smoothness of the evolution of the overall structure by tracking changes in the functions of the eigenvalues of the graph Laplacian, normalized Laplacian, or simply the adjacency matrix \cite{Donnat_2018}. A different perspective focused on the use of graph kernels to define similarities between graphs \cite{Jurman2015,Hammond2013,Scott2021} and the study of the product of states to evaluate the effect of initial conditions in information dynamics \cite{DomenicoPRE2020} and, dissimilarities in diffusive transport \cite{RiascosDiffusion2023}. To the best of our knowledge, a measure to compare synchronization or general nonlinear dynamics in systems coupled by networks is still lacking. A measure of this type is of central importance to the key theme of how these dynamics on networks relate to the structure more generally.
\\[2mm]
In this contribution, we introduce a general approach to compare two synchronization processes coupled through the information provided by a network. The paper is organized as follows. In Sec. \ref{Sec_Definitions}, we introduce general definitions and the method to detect differences in the dynamics of coupled oscillators on networks. We also present the Kuramoto model of identical oscillators and its linear approximation.  In Sec. \ref{Sec_Results} we apply the methods developed to the study of coupled systems that evolve with the Kuramoto model in different topologies. In particular, we explore analytically the effect of bias in a system of two coupled oscillators, modifications generated by couplings described by circulant matrices, and the effect of adding a new edge in a system initially coupled with a ring. We also compare the synchronization between all the nonisomorphic graphs with four nodes. Finally, we adapt all the methods developed to compare the Kuramoto model with its linear approximation in systems that evolve with the same coupling starting from random initial phases. We characterize the differences in the final configurations generated by each model using Shannon's entropy.  In Sec. \ref{Sec_Conclusions} we present the conclusions.  All the tests implemented to the measures of dissimilarity introduced in this research evidence their capacity to compare two synchronization processes in systems coupled by networks. The methods introduced can be applied to the study of a diverse variety of coupled oscillator systems,  and the entire approach provides insights into the information to be contemplated when comparing dynamical processes occurring in complex systems.
\section{General definitions}
\label{Sec_Definitions}
\subsection{Comparing two systems of coupled oscillators}
\label{Ref_Sec_Distance_Defs}
Let us consider a coupled set of oscillators on a connected network with $N$ nodes $i=1,\ldots ,N$. The phases $\theta_i(t)$ at each node evolve with continuous time $t$ starting at $t=0$. The structure of the network is described by an adjacency matrix $\mathbf{A}$ with elements $A_{ij}=1$ if there is a link connecting nodes $i$ and $j$ and $A_{ij}=0$ otherwise. The coupling between nodes is defined by a $N\times N$ matrix of weights $\mathbf{\Omega}$ with elements $\Omega_{ij}\geq 0$ for $i,j=1,\ldots,N$. The matrix $\mathbf{\Omega}$ is general in the sense that it can include the connectivity of a network as well as specific weights that define the dynamics. We are interested in the comparison of two synchronization processes defined by the coupling matrices $\mathbf{\Omega}$ and $\mathbf{\Omega}^\star$. The respective vector phases are denoted as $\vec{\theta}(t)$ and $\vec{\theta}^\star(t)$, with elements $\theta_i(t)$, $\theta_i^\star(t)$ for $i=1,2,\ldots, N$. 
\\[2mm]
In the study of synchronization processes, it is necessary to compare phases $\varphi$ taking values in the interval $[0,2\pi)$. In this manner, it is convenient to use a metric that compares directly two angles $0\leq \varphi_1,\varphi_2<2\pi$. We define the distance $d(\varphi_2-\varphi_1)$ as the shortest angle among $\varphi_1$ and $\varphi_2$, this is the difference between the angles module $2\pi$. In this manner $d(\varphi_2-\varphi_1)=d(\varphi_1-\varphi_2)$, $0 \leq d(\varphi_2-\varphi_1)\leq\pi$ and satisfy $\cos[d(\varphi_2-\varphi_1)]=\cos[\varphi_2-\varphi_1]$.
\\[2mm]
Then, in terms of the function $d$, we define the distance or {\it dissimilarity} $\mathcal{D}_s(t)$ between the dynamical processes with phases $\vec{\theta}(t)$ and $\vec{\theta}^\star(t)$ as
\begin{equation}\label{def_D_s}
	\mathcal{D}_s(t)\equiv \frac{1}{\pi}\sum_{\ell=1}^N d\left[\theta_\ell(t)-\theta^\star_\ell(t)\right],
\end{equation}
where the index $s$ denotes the modification of the initial condition in such a way $\theta_s(0)=\theta^\star_s(0)=\theta_0$ maintaining $\theta_\ell(0)=\theta^\star_\ell(0)=0$ for $\ell=1,\ldots,N$ and $\ell\neq s$. In the definition of $\mathcal{D}_s(t)$, the factor $\frac{1}{\pi}$ is included to express the global effect on the phases in terms of the maximum distance $\pi$ between two angles.
\\[2mm]
Therefore, $\mathcal{D}_s(t)$ in Eq. (\ref{def_D_s}) is similar to the total Manhattan distance between the phases of the processes $\vec{\theta}(t)$ and $\vec{\theta}^\star(t)$. The metric includes the information that the configurations of both systems at time $t$ are given by phases and lie on the surface of an $N$-dimensional hypertorus considering that at $t=0$, only the oscillator $s$ is dephased with an angle $\theta_0$. This initial condition is the same for both systems of coupled oscillators defined by $\mathbf{\Omega}$ and $\mathbf{\Omega}^\star$.
\\[2mm]
In addition, it is convenient to calculate the global value
\begin{equation}\label{Daver_general}
	\mathcal{\bar{D}}(t)\equiv \frac{1}{N}\sum_{s=1}^N \mathcal{D}_s(t)
\end{equation}
obtained considering independent modifications of the initial conditions for all the nodes $s=1,2,\ldots, N$. The values of $\mathcal{\bar{D}}(t)$ for $t$ large are denoted as $\mathcal{\bar{D}}_\infty\equiv \lim_{t\to \infty}\mathcal{\bar{D}}(t)$. This quantity helps us to compare the phases of two systems at large times and $\mathcal{\bar{D}}_\infty=0$ if these systems reach stationary states with the same phases.
\\[2mm]
Furthermore, the maximum global dissimilarity is defined as
\begin{equation}\label{Dmax_aver_def}
	\bar{\mathcal{D}}_{\mathrm{max}}\equiv \mathrm{max}\{\mathcal{\bar{D}}(t):t>0\}.
\end{equation}
In this manner, $\mathcal{\bar{D}}(t)$ captures globally how evolve the effect of having a non-null phase as an initial condition at each node $s$ of the network giving an average of the dissimilarities between two synchronization processes. In contrast,  $\mathcal{\bar{D}}_{\mathrm{max}}$ captures the differences with only one real value, but its calculation requires the evaluation of  $\mathcal{\bar{D}}(t)$ from $t=0$ to a time $t$ when the two processes reached stationary configurations of the phases.
\\[2mm]
The quantities introduced in this section allow characterizing the differences generated in the dynamics of two oscillator systems with different coupling matrices. The formalism presented is general and can be implemented to study oscillator systems defined by mathematical models that evolve deterministically, with parameters obtained from a probability distribution, or with stochastic differential equations (we refer the reader to the Refs. \cite{PikoBook,VespiBook,Arenas2008,J_Kurths_PhysRep2023} for a review of models describing synchronization processes). In Sec. \ref{Sec_Results}, we explore the methods developed in networked systems that evolve with the Kuramoto model and its linear approximation.
\subsection{Kuramoto model for identical oscillators}
The Kuramoto model is the paradigmatic approach to studying synchronization phenomena in nonlinear systems \cite{Rodrigues2016}. The oscillators in the original modeling are all-to-all coupled \cite{Yoshiki_K}; however, the coupling can be defined using the connectivity information of a network \cite{Arenas2008,RevModPhys_2005}.
\\[2mm]
In this section, we consider a system with $N$ nodes formed by identical  Kuramoto oscillators with a coupling structure defined by a network with a matrix of weights $\mathbf{\Omega}$, where the natural frequency $\omega$ is the same for each oscillator \cite{Taylor_2012}. However, all the methods developed can be adapted to the analysis of systems where the oscillators have different characteristics; for example, in systems with natural frequencies sampled from a probability distribution. At a time $t$, the oscillators are characterized by their phases $\theta_i(t)$ with $i=1,2,\ldots,N$ \cite{Arenas2008,RevModPhys_2005,Rodrigues2016}.  In this system,  the evolution of phases of identical Kuramoto oscillators placed in the nodes is described by the set of nonlinear coupled differential equations  \cite{Taylor_2012}
\begin{equation}\label{normalkuramoto}
 	\frac{d\theta_i(t)}{dt}=\omega_i+\lambda\sum_{j=1}^{N}\Omega_{ij}\sin[\theta_j(t)-\theta_i(t)],
\end{equation}
where $\lambda>0$ corresponds to the global coupling strength of the system and $\omega_i$ denotes the natural frequency of the $i$-th oscillator. For identical oscillators $\omega_i=\omega$;  in addition, the system in Eq. (\ref{normalkuramoto}) satisfies rotational symmetry being invariant under the transformation $\theta_i\to\theta_i+\omega t$. Then, it is possible to set $\omega=0$ and rescale the time by setting $\lambda=1$ \cite{Townsend}. In this manner, Eq. (\ref{normalkuramoto})  is transformed into \cite{Taylor_2012,Townsend,Jadbabaie_Motee_Barahona_2004}
\begin{equation}\label{kuramoto_omega}
    \frac{d\theta_i(t)}{dt}=\sum_{j=1}^{N
    }\Omega_{ij}\sin[\theta_j(t)-\theta_i(t)]
\end{equation}
for $i=1,2,\ldots,N$. The dynamics defined by Eq. (\ref{kuramoto_omega}) is a generalization of the model introduced by Kuramoto in Ref. \cite{Yoshiki_K}
to the case of identical oscillators in weighted networks (we refer the reader to Refs.  \cite{Arenas2008,Rodrigues2016,J_Kurths_PhysRep2023} for detailed discussions of the Kuramoto model on networks). 
\\[2mm]
One of the main features of the system in Eq. (\ref{kuramoto_omega}) is that in some cases, the phases evolve to reach a global synchronized state. The conditions under a system of Kuramoto identical oscillators reach complete synchronization are still under study; however, the topology of the network describing the coupling plays an important role \cite{Taylor_2012,Townsend,Ling, HA20101692}. A common quantity to measure the phase coherence of the oscillators is through the macroscopic order parameter $r(t)$ defined by \cite{Arenas2008}
\begin{equation}\label{orderparam}
    r(t)=\dfrac{1}{N}\left|\sum_{j=1}^{N}  \exp\left[\mathbf{i}\theta_j(t)\right]\right|,
\end{equation}
where $\mathbf{i}=\sqrt{-1}$. From the definition in Eq. (\ref{orderparam}), $0 \leq r(t) \leq 1$. In the case of complete phase coherence $r(t) = 1$, whereas $r(t) = 0$ for completely incoherent oscillators.
\\[2mm]
In the particular case of symmetric coupling matrices, i.e., with $\Omega_{ij}=\Omega_{ji}$, the sum of the phases in Eq. (\ref{kuramoto_omega}) allows obtaining $\frac{d}{dt} \sum_{i=1}^N \theta_i(t)=0$; therefore, for symmetric $\mathbf{\Omega}$
\begin{equation}
	\frac{1}{N}\sum_{i=1}^N \theta_i(t)= \frac{1}{N}\sum_{i=1}^N \theta_i(0),
\end{equation}
showing that, in this particular case, the average phases are a constant in the dynamics.

\subsection{Linear approximation: Laplacian matrix of weighted networks}
\label{SubSec_Linear}
In the Kuramoto model, for some initial conditions closer to the synchronization is valid the linearization of the dynamics [using the approximation $\sin(x)\approx x$ for $x$ small in Eq. (\ref{kuramoto_omega})].
In this approximation of the Kuramoto model of identical oscillators, the dynamics of the synchronous state can be assessed using a  master stability function approach. In this section, we briefly describe the linear approximation of the Kuramoto model. For small values of $\theta_j(t)-\theta_i(t)$, is valid the linear approximation of Eq. (\ref{kuramoto_omega})
\begin{align}\nonumber
    \frac{d\theta_i(t)}{dt}&\approx\sum_{j=1}^{N
    }\Omega_{ij}[\theta_j(t)-\theta_i(t)]\\
    &=-\sum_{j=1}^N (\mathcal{K}_i\delta_{ij}-\Omega_{ij})\theta_j(t).
\label{linear_approx1}
\end{align}
Here the generalized degree $\mathcal{K}_i$ of the node $i$ is defined as  $\mathcal{K}_i\equiv\sum_{\ell=1}^N \Omega_{i\ell}$ and $\delta_{ij}$ denotes the Kronecker delta. In this manner, we define the Laplacian matrix $\mathbf{L}$ of a weighted network, with elements $L_{ij}$ given by
\begin{equation}\label{Laplacian_weighted}
    L_{ij}\equiv\mathcal{K}_i\delta_{ij}-\Omega_{ij}.
\end{equation}
%
Then, the linear approximation in Eq. (\ref{linear_approx1}) defines the dynamical process
\begin{equation}\label{dtheta_linear}
\frac{d\theta_i(t)}{dt}=-\sum_{j=1}^N L_{ij}\theta_j(t).
\end{equation}
The solution of this set of linear coupled differential equations allows to obtain
\begin{equation}\label{linear_exp}
\theta_i(t)=\sum_{j=1}^N\left(e^{-t \mathbf{L}}\right)_{ij}\theta_j(0).
\end{equation}
Now, using Dirac's notation for the eigenvectors, we have a set of right eigenvectors $\{\left|\Psi_j\right\rangle\}_{j=1}^N$  that satisfy the eigenvalue equation
$\mathbf{L}\left|\Psi_j\right\rangle=\mu_j\left|\Psi_j\right\rangle$ for $j=1,\ldots,N$. With this information, we define the matrix $\mathbf{Q}$  with elements $Q_{ij}=\left\langle i|\Psi_j\right\rangle$  and the diagonal matrix $\mathbf{\Lambda}(t)=\textrm{diag}(e^{-t\mu_1},e^{-t\mu_2},\ldots,e^{-t\mu_N})$. These matrices satisfy $e^{-t \mathbf{L}}=\mathbf{Q}\mathbf{\Lambda}(t)\mathbf{Q}^{-1}$, 
where $\mathbf{Q}^{-1}$ is the inverse of $\mathbf{Q}$. Using the matrix $\mathbf{Q}^{-1}$, we define the set of left eigenvectors $\{\left\langle \bar{\Psi}_i\right|\}_{i=1}^N$ with components $\left\langle \bar{\Psi}_i|j\right\rangle=(\mathbf{Q}^{-1})_{ij}$ (see Refs. \cite{FractionalBook2019,DirectedFractional_PRE2020,RiascosDiffusion2023} for a detailed discussion of the Laplacian matrix for undirected, directed and weighted networks). Therefore, the solution for the linear dynamics in  Eq. (\ref{linear_exp}) takes the form
\begin{equation}\label{theta_i_eigensys}
    \theta_i(t)=\sum_{j=1}^N\sum_{\ell=1}^N e^{-t\mu_\ell}\langle i|\Psi_\ell\rangle \langle \bar{\Psi}_\ell|j\rangle \theta_j(0).
\end{equation}
In this manner, Eq. (\ref{theta_i_eigensys}) allows obtaining analytically the phases of oscillators in terms of the eigenvalues $\mu_\ell$ and the sets of eigenvectors $\left|\Psi_\ell\right\rangle$ and $\left\langle \bar{\Psi}_\ell\right|$ of the matrix $\mathbf{L}$ for the linear approximation of the Kuramoto model in Eq. (\ref{dtheta_linear}).
\section{Results}
\label{Sec_Results}
In this section we apply the approach presented in Sec. \ref{Sec_Definitions} to diverse cases. We analyze the effect of biased couplings in a system with two oscillators, the evolution of systems with couplings described by circulant matrices, the effect of adding a new edge on a ring graph, and the comparison of the synchronization for all the connected networks with $N=4$ nodes. Also, we study the differences between the Kuramoto model and its linear approximation in four structures including deterministic and random networks.

\begin{figure*}[t]
	\centering
	\includegraphics*[width=1.0\textwidth]{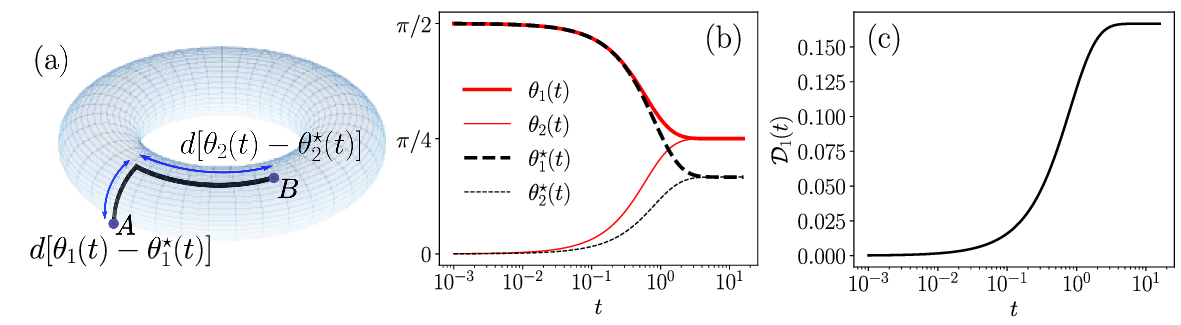}
	\vspace{-2mm}
	\caption{\label{Fig_1} Dissimilarity in the synchronization of two coupled oscillators.  (a) Two dimensional torus illustrating the phase separations $d[\theta_1(t)-\theta^\star_1(t)]$ and $d[\theta_2(t)-\theta^\star_2(t)]$ between the point $A$ with phases $[\theta_1(t),\theta_2(t)]$ and $B$ with phases $[\theta^\star_1(t),\theta^\star_2(t)]$. (b) Phases $\theta_1(t)$ and $\theta_2(t)$ for the Kuramoto model with symmetric coupling and the modified dynamics $\theta^\star_1(t)$ and $\theta^\star_2(t)$ with a value $\alpha=1/2$, the results are obtained from the analytical solution in Eqs. (\ref{sol_coupledN2}) and (\ref{varphi_N2}) with initial conditions $\theta_1(0)=\theta_1^\star(0)=\pi/2$ and $\theta_2(0)=\theta_2^\star(0)=0$.   (c)  Dissimilarity $\mathcal{D}_1(t)$ as defined in Eq. (\ref{def_D_s}) for $N=2$ evaluated with the phases showed in panel (b).}
\end{figure*}

\subsection{Two oscillators: effect of biased coupling}
As a first example, let us analyze a system with $N=2$ Kuramoto oscillators with phases $\phi_1(t)$, $\phi_2(t)$ defined by
\begin{subequations}\label{KM_coupledN2}
\begin{align}
\frac{d\phi_1(t)}{dt}&=\sin\left[\phi_2(t)-\phi_1(t)\right],\\
\frac{d\phi_2(t)}{dt}&=\alpha\,\sin\left[\phi_1(t)-\phi_2(t)\right]
\end{align}
\end{subequations}
with the initial condition $\phi_1(0)=\phi_0$ and $\phi_2(0)=0$ and $\alpha>0$ a real parameter.
\\[2mm]
The general set of Eqs. (\ref{KM_coupledN2}) can be solved analytically considering the variable $u(t)=\phi_1(t)-\phi_2(t)$. The obtained solution is given by
\begin{subequations}\label{sol_coupledN2}
\begin{align}
\phi_1(t)&=\frac{1}{1+\alpha}\left[\alpha\,\phi_0+2\tan^{-1}\varphi_\alpha(t,\phi_0)\right],\\
\phi_2(t)&=\frac{\alpha}{1+\alpha}\left[\phi_0-2\tan^{-1}\varphi_\alpha(t,\phi_0)\right],
\end{align}
\end{subequations}
valid for $0\leq \phi_0<\pi$ and where
\begin{equation}\label{varphi_N2}
\varphi_\alpha(t,\phi_0)\equiv \tan\left(\frac{\phi_0}{2}\right)\exp[-(1+\alpha)t].
\end{equation}
Once presented the analytical solution for the general Kuramoto model in Eqs. (\ref{sol_coupledN2}) and (\ref{varphi_N2}), let us consider two particular dynamical systems reaching complete synchronization. The first one, with phases $\theta_1(t)$ and $\theta_2(t)$ is the system in Eq. (\ref{KM_coupledN2}) with $\alpha=1$. For this unbiased case, the coupling matrix $\mathbf{\Omega}$ is 
\begin{equation}
\mathbf{\Omega}=\left(\begin{array}{cc}
   0  & 1 \\
   1  & 0
\end{array}\right),
\end{equation}
the dynamics of $\theta_1(t)$ and $\theta_2(t)$ is considered as reference. In a similar manner, the second process, including the value $\alpha$ in Eq. (\ref{KM_coupledN2}) is defined by the phases $\theta^\star_1(t)$ and $\theta^\star_2(t)$ and the coupling matrix $\mathbf{\Omega}^\star$ 
\begin{equation}\label{Omega_star_N2}
\mathbf{\Omega}^\star=\left(\begin{array}{cc}
   0  & 1 \\
   \alpha  & 0
\end{array}\right).
\end{equation}
In both processes, we consider the same initial conditions $\theta_1(0)=\theta^\star_1(0)=\theta_0$ and $\theta_2(0)=\theta^\star_2(0)=0$, with  $0\leq \theta_0<\pi$. Here, it is worth noticing that both systems reach synchronization. According to Eq. (\ref{sol_coupledN2}), for the first (reference) system, in the limit $t\to\infty$
\begin{equation}\label{Limit_N2symmetric}
\theta_1(t\to \infty)=\theta_2(t\to \infty)=\frac{\theta_0}{2}.
\end{equation}
Similarly, for the second (modified) system
\begin{equation}\label{Limit_N2alpha}
\theta_1^\star(t\to \infty)=\theta_2^\star(t\to \infty)=\frac{\alpha}{\alpha+1}\theta_0.
\end{equation}
The results in Eqs. (\ref{Limit_N2symmetric}) and (\ref{Limit_N2alpha}) are also obtained using the linear approximation discussed in Sec. \ref{SubSec_Linear} with the Laplacian matrices associated with the coupling matrices   $\mathbf{\Omega}$ and $\mathbf{\Omega}^\star$.
\\[2mm]
In Fig. \ref{Fig_1} we explore the phases of the reference process $\vec{\theta}(t)\equiv[\theta_1(t),\theta_2(t)]$ and the modified dynamics $\vec{\theta}^\star(t)\equiv [\theta_1^\star(t),\theta_2^\star(t)]$ to calculate $\mathcal{D}_1(t)$ given by Eq. (\ref{def_D_s}). 
In Fig. \ref{Fig_1}(a), we illustrate geometrically on a torus surface the metric implemented. In this example, each configuration of the system is represented by a point on the surface. Then, the value $d[\theta_1(t)-\theta^\star_1(t)]+d[\theta_2(t)-\theta^\star_2(t)]$ quantifies the separation of the configurations in $\vec{\theta}(t)$ and $\vec{\theta}^\star(t)$, this value (divided by $\pi$) is used to calculate $\mathcal{D}_1(t)$. 
\\[2mm]
In Fig. \ref{Fig_1}(b), we present the numerical evaluation of $\theta_1(t)$, $\theta_2(t)$ for the symmetric coupling used as a reference. The values are obtained from the analytical result in Eq. (\ref{sol_coupledN2}) with $\alpha=1$, we also show the modified dynamics $\theta_1^\star(t)$, $\theta_2^\star(t)$ with $\alpha=1/2$. In both cases, we use the value $\theta_0=\pi/2$ associated with the modification of the initial phase of the node $s=1$. The results reveal the evolution of the phases to reach synchronized configurations and how the second process is modified as a consequence of the asymmetry generated by $\alpha$. The results for $t\gg 1 $ also coincide with the values $\theta_1^\star(t\to \infty)=\theta_2^\star(t\to \infty)=\pi/4$ and $\theta_1^\star(t\to \infty)=\theta_2^\star(t\to \infty)=\pi/6$ in Eqs. (\ref{Limit_N2symmetric}) and (\ref{Limit_N2alpha}).
\\[2mm]
The numerical values for the phases obtained in Fig.  \ref{Fig_1}(b) allow the evaluation of $\mathcal{D}_1(t)$ in Eq. (\ref{def_D_s}) since we are using the initial conditions $\theta_1(0)=\theta_1^\star(0)=\pi/2$ and $\theta_2(0)=\theta_2^\star(0)=0$, where only the initial phase of the node $s=1$ is modified with an angle $\theta_0=  \pi/2$. $\mathcal{D}_1(t)$ quantifies the differences in the synchronization generated in the two processes maintaining the same initial condition; the numerical results are presented in Fig. \ref{Fig_1}(c). In particular, for $t\gg 1$, we have
\begin{equation*}
\mathcal{D}_1(t\to \infty)=\frac{1}{\pi}\left[d\left(\frac{\pi}{4}-\frac{\pi}{6}\right)+
d\left(\frac{\pi}{4}-\frac{\pi}{6}\right)
\right]=\frac{1}{6},
\end{equation*}
a result that coincides with the numerical results for $t>5$. In general, the form of $\mathcal{D}_1(t)$ evidence the modifications in the synchronization starting from the same initial condition $\mathcal{D}_1(t\to 0)\to 0$  and evolving to $\mathcal{D}_1(t\to \infty)>0$ showing that the synchronized states are different as a direct consequence of the asymmetry in the coupling defined by the matrix $\mathbf{\Omega}^\star$ in Eq. (\ref{Omega_star_N2}).

\begin{figure*}[t]
	\centering
	\includegraphics*[width=0.95\textwidth]{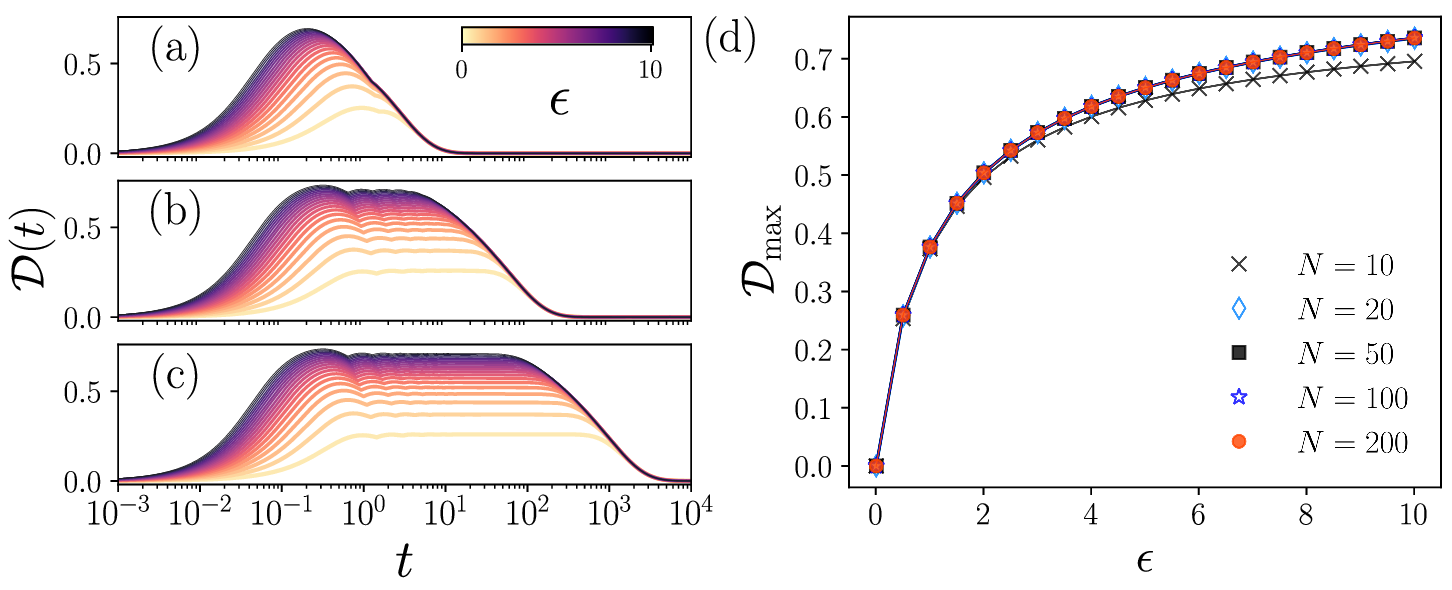}
	\vspace{-2mm}
	\caption{\label{Fig_2} Phase differences between the synchronization on a ring and on weighted interacting cycles. $\mathcal{D}(t)$ as a function of $t$ for networks with sizes: (a) $N=10$, (b) $N=50$, (c) $N=200$, the weights $\epsilon$ are codified in the colorbar. The results are obtained with the numerical evaluation of Eq. (\ref{def_D_circulant}). (d) $\mathcal{D}_{\mathrm{max}}$ in Eq. (\ref{MaxD_circulant}) as a function of $\epsilon$ for networks with different sizes $N$.}
\end{figure*}

\subsection{Systems coupled with circulant matrices}
In this section, we apply the general approach introduced in Sec. \ref{Ref_Sec_Distance_Defs} to the study of processes coupled using circulant matrices. The results are illustrated with the analysis of the evolution of the Kuramoto model in Eq. (\ref{kuramoto_omega}) on interacting cycles and biased rings.
\\[2mm]
A circulant matrix $\mathbf{C}$ is a  $N \times N$ matrix defined by entries denoted as $C_{ij}$ \cite{Gray_Review2006,VanMieghem2011}. Each column has real elements $c_0, c_1,\ldots,c_{N-1}$ ordered in such a way that $ c_0 $ describes the diagonal elements and $C_{ij}=c_{(i-j)\text{mod}\, N}$. In this manner \cite{VanMieghem2011}
\begin{equation} \label{matC}
	\mathbf{C}=
	\left(
	\begin{array}{ccccc}
		c_0 & c_{N-1} & c_{N-2} & \ldots & c_1\\
		c_1 & c_{0} & c_{N-1} & \ldots & c_2\\
		c_2 & c_{1} & c_{0} & \ldots & c_3\\
		\vdots & \vdots & \vdots & \ddots & \vdots \\
		c_{N-1} & c_{N-2} & c_{N-3} &\ldots & c_0\\
	\end{array}\right) \, .
\end{equation}
\\[2mm]
The comparison between two synchronization dynamical processes with coupling matrices $\mathbf{\Omega}$ and $\mathbf{\Omega}^\star$ is given by
\begin{equation}\label{def_D_circulant}
\mathcal{D}(t)\equiv \frac{1}{\pi}\sum_{\ell=1}^N d\left[\theta_\ell(t)-\theta^\star_\ell(t)\right].
\end{equation}
In this case, the initial condition is  $\theta_s(0)=\theta^\star_s(0)=\theta_0$ maintaining $\theta_\ell(0)=\theta^\star_\ell(0)=0$ for $\ell=1,\ldots,N$ and $\ell\neq s$. However, due to the regularity of both coupling structures  describing the matrices $\mathbf{\Omega}$ and $\mathbf{\Omega}^\star$, all the results for $s=1,2,\ldots,N$ are equivalent. As a consequence, Eq. (\ref{Daver_general}) takes the form
\begin{equation}
\mathcal{D}(t)=\bar{\mathcal{D}}(t),
\end{equation}
a result valid when the couplings in the two processes are described by circulant matrices. Also, it is convenient to define $\mathcal{D}_\infty\equiv \lim_{t\to \infty} \mathcal{D}(t)$  and, the maximum separation $\mathcal{D}_{\mathrm{max}}$ as
\begin{equation}\label{MaxD_circulant}
\mathcal{D}_{\mathrm{max}}\equiv\max\{\mathcal{D}(t)\,\forall\, t\geq 0\}.
\end{equation}
\subsubsection{Interacting cycles}
\label{subsec_intcycles}
Let us now explore the differences generated in the synchronization when a set of weighted edges is added to a ring. In this manner, we consider a ring with $N$ nodes as the reference process with a coupling matrix $\mathbf{\Omega}$ defined by a circulant matrix [as in Eq. (\ref{matC})] with non-null entries $c_1=c_{N-1}=1$. Then, each node interacts only with two nearest neighbors forming a cyclic structure. The second process is described by a coupling matrix  $\mathbf{\Omega}^\star$ considering a ring (the original matrix  $\mathbf{\Omega}$) and including additional links with a weight $\epsilon \geq 0$ connecting all the nodes at distance $2$ in the original ring. $\mathbf{\Omega}^\star$ is a circulant matrix defined by the non-null entries $c_1=c_{N-1}=1$ and $c_2=c_{N-2}=\epsilon$.
\\[2mm]
From the evolution of synchronization in the systems with coupling $\mathbf{\Omega}$ and $\mathbf{\Omega}^\star$, we can study the effect of $\epsilon$ in the global dynamics. In Fig. \ref{Fig_2}, we present the numerical values of $\mathcal{D}(t)$ and $\mathcal{D}_{\mathrm{max}}$ for different weights $\epsilon$ and  sizes $N$. In Figs. \ref{Fig_2}(a) to \ref{Fig_2}(c) we explore $\mathcal{D}(t)$ as a function of $t$ for networks with $N=10,\, 50,\, 200$, the values are depicted using different colors for $0<\epsilon\leq 10$ codified in the colorbar. The results are obtained with the numerical solution of Eq. (\ref{kuramoto_omega}) for the couplings $\mathbf{\Omega}$ and $\mathbf{\Omega}^\star$, with the initial condition $\theta_1(0)=\theta_1^\star(0)=\theta_0=\pi/2$ and $\theta_\ell(0)=\theta_\ell^\star(0)=0$ for $\ell=2,3,\ldots,N$. With this information, we calculate  $\mathcal{D}(t)$ in Eq. (\ref{def_D_circulant}), due to the regularity of the coupling networks, the initial dephase in any node is sufficient to determine the global response $\mathcal{D}(t)$. The results for $\mathcal{D}(t)$ show how initially (at $t\to 0$), the two processes are similar and the differences generated by $\epsilon$ are observed after $t\geq 0.01$ independently of the network size. The values increase until reaching a plateau that depends on the size of the network to then decay to  $\mathcal{D}_\infty=0$ for large times showing that the synchronized configurations are the same for the two dynamics. This particular limit is a consequence of the symmetry of both coupling matrices  $\mathbf{\Omega}$ and $\mathbf{\Omega}^\star$.
\\[2mm]
The results in Figs. \ref{Fig_2}(a) to \ref{Fig_2}(c) also showing that the maximum value of $\mathcal{D}(t)$ is a good measure to quantify the global effect of $\epsilon$. In  Fig. \ref{Fig_2}(d) we present  $\mathcal{D}_{\mathrm{max}}$ as defined in Eq. (\ref{MaxD_circulant}) for different $\epsilon$ and $N$. The numerical values reveal how $\mathcal{D}_{\mathrm{max}}$ increases monotonically with $\epsilon$ for each $N$, whereas for $\epsilon=0$,  $\mathcal{D}_{\mathrm{max}}=0$ since, in this particular limit, both coupling matrices coincide. Also, it is important to notice that, for $N\geq 20$, the  curves for $\mathcal{D}_{\mathrm{max}}$ are similar, only some small deviations are observed for the networks with $N=10$.
\\[2mm]
All the results presented in Fig. \ref{Fig_2} evidence that  $\mathcal{D}(t)$ in Eq. (\ref{def_D_circulant}) provides a good measure to evaluate the differences generated by the weights described by $\epsilon$ in the extra edges incorporated in the modified synchronization process. It is also important to highlight that, for $t$ large, the final synchronized phases are the same in the processes with $\mathbf{\Omega}$ and $\mathbf{\Omega}^\star$, something that becomes evident with the limit $\mathcal{D}_\infty=0$. In this manner, all the differences observed using $\mathcal{D}_{\mathrm{max}}$ are due to transitory states of the systems where the phases produce $\mathcal{D}(t)>0$. The duration of this transient interval depends to a greater extent on the size of the network.
\subsubsection{Biased coupling on rings}
\begin{figure*}[t]
	\centering
	\includegraphics*[width=1.0\textwidth]{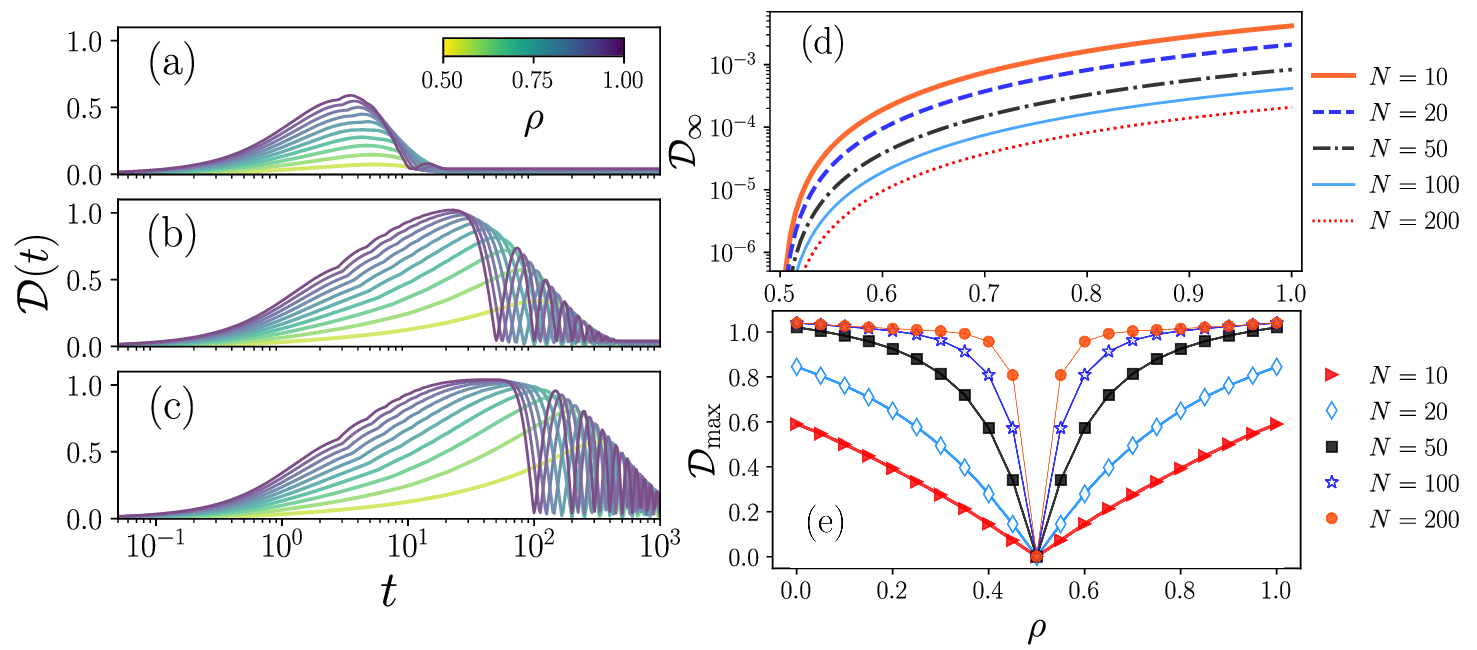}
	\vspace{-7mm}
	\caption{\label{Fig_3} Phase differences between the synchronization on a ring and on  rings with bias $\rho$. $\mathcal{D}(t)$ as a function of $t$ for networks with sizes: (a) $N=10$, (b) $N=50$, (c) $N=100$, the weights $\rho$ are codified in the colorbar. The results are obtained with the numerical evaluation of Eq. (\ref{def_D_circulant}). (d) $\mathcal{D}_{\infty}=\lim_{t\to\infty} \mathcal{D}(t)$ and (e) $\mathcal{D}_{\mathrm{max}}$ in Eq. (\ref{MaxD_circulant}) as a function of $\rho$ for rings with sizes $N=10,20,50,100,200$.}
\end{figure*}
In this section, we explore the effect of bias in the coupling of rings. The approach is similar to the implemented for interacting cycles in Sec. \ref{subsec_intcycles}. We use as reference the synchronization on a ring with $N$ nodes described by a coupling matrix $\mathbf{\Omega}$ defined by a circulant matrix with non-null entries $c_1=c_{N-1}=1/2$. The second process is determined by the circulant matrix $\mathbf{\Omega}^\star$ with non-null entries $c_1=\rho$ and $c_{N-1}=1-\rho$ where the parameter $\rho$ fulfills $0\leq \rho \leq 1$. In this way, $\mathbf{\Omega}^\star$ incorporates the effect of having a bias in the coupling, recovering the process used as reference when $\rho=1/2$.
\\[2mm]
In Fig. \ref{Fig_3} we present $\mathcal{D}(t)$, $\mathcal{D}_\infty$ and $\mathcal{D}_{\mathrm{max}}$ for different $\rho$ and sizes $N$.  Our findings require the numerical solution of Eq. (\ref{kuramoto_omega}) for the couplings $\mathbf{\Omega}$ and $\mathbf{\Omega}^\star$. The values  $\mathcal{D}(t)$  are generated from Eq. (\ref{def_D_circulant}) with the initial condition $\theta_1(0)=\theta_1^\star(0)=\theta_0=\pi/2$ and $\theta_\ell(0)=\theta_\ell^\star(0)=0$ for $\ell=2,3,\ldots,N$. In Figs. \ref{Fig_3}(a) to \ref{Fig_3}(c) we explore $\mathcal{D}(t)$ as a function of $t$ for values of the bias $0.5<\rho\leq 1$ codified in the colorbar. In the case with $N=10$ in Fig.  \ref{Fig_3}(a), we see how $\mathcal{D}(t)$ grows with $t$ until reaching a maximum and decaying, in the interval $t>20$, $\mathcal{D}(t)$ is constant with $\mathcal{D}(t)>0$. In networks with $N=50,\, 100$ [in Figs. \ref{Fig_3}(b) and \ref{Fig_3}(c)] $\mathcal{D}(t)$ grows with $t$; however, after reaching the maximum it is clear an oscillatory behavior that gradually decays in its amplitude. The results also show that for $t$ large, $\mathcal{D}(t)$ is non-null as a consequence of the asymmetry in the couplings, we explore this feature in Fig. \ref{Fig_3}(d) where is plotted $\mathcal{D}_\infty$ as a function of $\rho$ for $0.5<\rho\leq 1$.  Additionally, $\mathcal{D}_{\mathrm{max}}$ is a first approximation to capture with just one number the effect of the bias. Our findings are shown in Fig. \ref{Fig_3}(e), presenting $\mathcal{D}_{\mathrm{max}}$ as a function of $\rho$ in the interval $0\leq\rho\leq 1$. We see that $\mathcal{D}_{\mathrm{max}}$ increase monotonically with $|\rho-1/2|$ and $\mathcal{D}_{\mathrm{max}}=0$ for $\rho=1/2$; also, the values $\mathcal{D}_{\mathrm{max}}$ increase with the network size.
\\[2mm]
To gain some intuition of the oscillatory behavior observed in Figs. \ref{Fig_3}(b) to \ref{Fig_3}(c), we explore the linear approximation of the Kuramoto model presented in Sec. \ref{SubSec_Linear}. In this manner, from Eq. (\ref{Laplacian_weighted}), for the process used as reference the Laplacian matrix $\mathbf{L}$ is circulant with non-null elements $c_0=1$, $c_1=c_{N-1}=-1/2$. For the biased processes, the Laplacian matrix $\mathbf{L}^\star$ is defined by the non-null entries $c_0=1$, $c_1=-\rho$, $c_{N-1}=-(1-\rho)$. Here, it is important to mention that, strictly speaking,  $\mathbf{L}$ and  $\mathbf{L}^\star$ are the normalized Laplacians associated with continuous-time random walks on networks (see details in Ref. \cite{RiascosDiffusion2023}).
\\[2mm]
Also, it can be shown that the eigenvalues $\xi_{l}$ (with  $l=1,2,\ldots,N$) of $\mathbf{L}$ are \cite{RiascosDiffusion2023} 
\begin{equation}\label{eigen_ring}
	\xi_{l}=1-\cos\left[\frac{2\pi}{N}(l-1)\right].
\end{equation}
For the second process, the eigenvalues $\xi^\prime_{l}$ of $\mathbf{L}^\star$ are \cite{RiascosDiffusion2023}
\begin{multline}\label{eigen_ring_biased}
	\xi^\prime_{l}=1-(1-\rho)\exp\left[\mathbf{i}\frac{2\pi}{N}(l-1)\right]\\
	-\rho\exp\left[-\mathbf{i}\frac{2\pi}{N}(l-1)\right].
\end{multline}
In particular, denoting $\varphi_l\equiv\frac{2\pi}{N}(l-1)$, we rewrite Eq. (\ref{eigen_ring_biased}) to have
\begin{equation}\label{eigen_ring_biased2}
	\xi^\prime_{l}=\xi_l+\mathbf{i}(2\rho-1)\sin\varphi_l.
\end{equation}
\begin{figure*}[t]
	\centering
	\includegraphics*[width=0.95\textwidth]{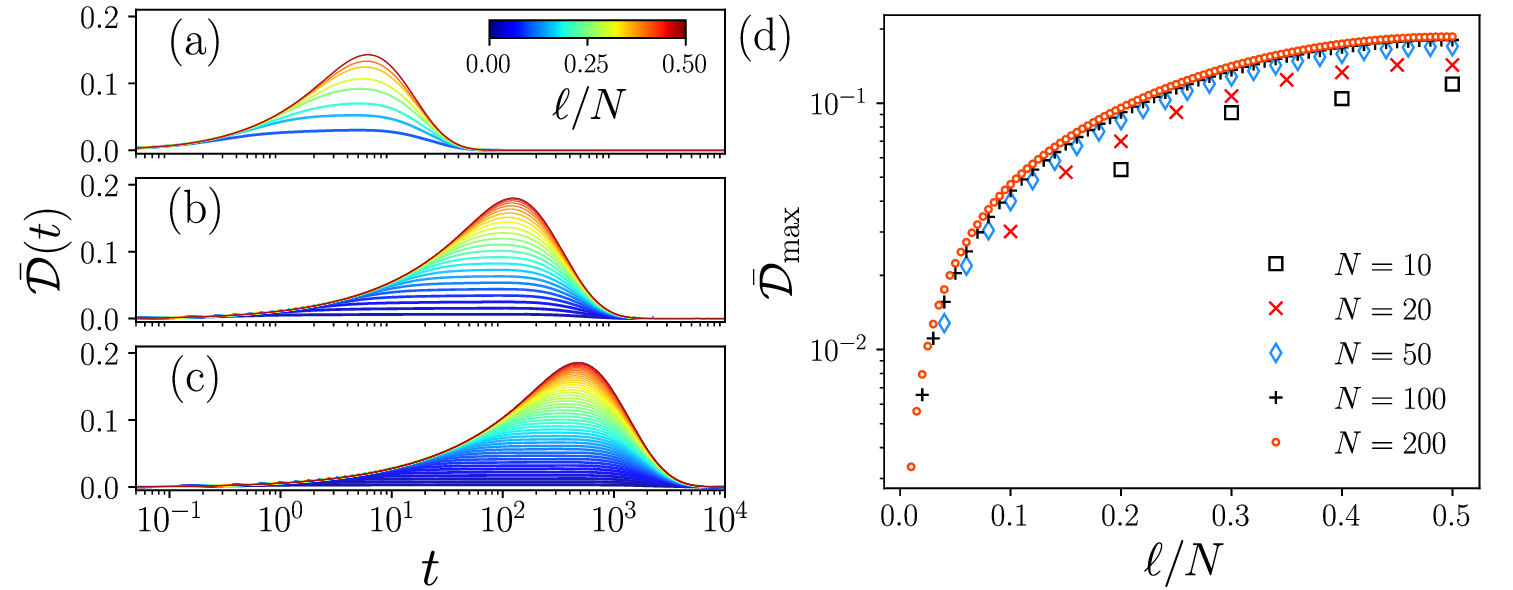}
	\vspace{-2mm}
	\caption{\label{Fig_4} 
		Dissimilarity between the synchronization on a ring and on a ring with an additional edge connecting two nodes at a distance $\ell$ in the original ring. $\bar{\mathcal{D}}(t)$ as a function of $t$ for different values of $\ell=2,3,\ldots,\lfloor N/2\rfloor$, codified in the colorbar. The results are obtained with the numerical evaluation of Eq.  (\ref{Daver_general}) for networks with: (a) $N=20$, (b) $N=100$, (c) $N=200$. (d)  $\bar{\mathcal{D}}_{\mathrm{max}}$ in Eq. (\ref{Dmax_aver_def}) as a function of $\ell/N$ for networks with different size $N$  (see details in the main text). 
	}
\end{figure*}
The value $\Delta_l\equiv(2\rho-1)\sin\varphi_l$ quantifies the modifications in the eigenvalues $\xi_l$ of $\mathbf{L}$ in Eq. (\ref{eigen_ring}). In addition, for circulant matrices, the right eigenvectors $\{|\Psi_m\rangle\}_{m=1}^{N}$ have the components $\langle l |\Psi_m\rangle=\frac{1}{\sqrt{N}}e^{-\mathbf{i}\frac{2\pi}{N}(l-1)(m-1)}$ and $\langle\Psi_l|m\rangle=\frac{1}{\sqrt{N}}e^{\mathbf{i}\frac{2\pi}{N}(l-1)(m-1)}$ \cite{VanMieghem2011}. Therefore, using the result in Eq. (\ref{theta_i_eigensys}), we have for the reference process
\begin{equation}
	\theta_i(t)=\frac{\theta_0}{N}\sum_{l=1}^N e^{-\xi_l t}e^{-\mathbf{i}\varphi_l(i-1)},
\end{equation}
whereas for the synchronization in the system with bias
\begin{equation}
	\theta_i^\star(t)=\frac{\theta_0}{N}\sum_{l=1}^N e^{-\xi^\prime_l t}e^{-\mathbf{i}\varphi_l(i-1)}.
\end{equation}
Therefore:
\begin{align}\nonumber
	\theta_i(t)-\theta_i^\star(t)&=\frac{\theta_0}{N}\sum_{l=1}^Ne^{-\xi_l t}e^{-\mathbf{i}\varphi_l(i-1)}\left(1-e^{-\mathbf{i}\Delta_l t}\right)\\ \label{Delta_Theta_rho1}
 &=\frac{\theta_0}{N}\sum_{l=1}^Ne^{-\xi_l t}g_{il}(t,\rho)
\end{align}
where 
\begin{multline}\label{Delta_Theta_rho2}
   g_{il}(t,\rho)\equiv \cos[\varphi_l(i-1)]\\-\cos[\varphi_l(i-1)+t\,(2\rho-1)\sin\varphi_l]. 
\end{multline}
The result in Eqs. (\ref{Delta_Theta_rho1}) and (\ref{Delta_Theta_rho2}) shows how the differences in the phases $\theta_i(t)-\theta_i^\star(t)$ in the linear approximation have an oscillatory behavior with characteristic frequencies proportional to $2\rho-1$. Oscillations of this type are also present in the solution of the Kuramoto model evidenced in the results for $\mathcal{D}(t)$ in Fig. \ref{Fig_3}. 
\\[2mm]
Our findings in this section show that the effect of bias on the coupling produces different behaviors from what we found for the addition of links with weights explored in Sec. \ref{subsec_intcycles}. For the case with bias highlights the oscillatory behavior of $\mathcal{D}(t)$, this effect is marked in larger networks. In addition, in the limit of large times  $\mathcal{D}_\infty>0$, both features are because the $\mathbf{\Omega}^\star$ matrix is not symmetric. Finally, it is also important to see that $\mathcal{D}_{\mathrm{max}}$ is associated with differences in the transitory states and depends on the size of the network, showing that the global effects of the bias are very different for the diverse sizes explored.
\subsection{Effect of a chord on a ring}
In this section we illustrate the effect of adding a new edge that breaks the symmetry of a cyclic structure. To this end, we compare the dynamics generated by $\mathbf{\Omega}$ for a ring with $N$ nodes defined by the adjacency matrix, a circulant matrix as in Eq. (\ref{matC}) with non-null elements $c_1=c_{N-1}=1$. The second process generated by $\mathbf{\Omega}^\star$ describes the synchronization dynamics on a modified network defined by a ring with an additional edge connecting two nodes at a distance $\ell=2,3,\ldots,\lfloor N/2\rfloor$ in the original ring. In the graph theory literature this type of edge is called a chord \cite{west2001graph_theory}. In $\mathbf{\Omega}^\star$ all the entries associated with the edges have a value 1 and 0 in other cases.
\\[2mm]
We explore the effect of $\ell$ when we compare the synchronization in the ring and the ring with a chord. Here, it is worth mentioning that in all the cases the matrices $\mathbf{\Omega}$ and $\mathbf{\Omega}^\star$ only differ in two particular entries. Furthermore, for any $\ell$, $\sum_{i,j}(\Omega^\star_{ij}-\Omega_{ij})=2$ quantifying the two non-null entries associated to the chord and revealing that the direct comparison of the matrical elements does not capture the differences between the dynamics generated by $\mathbf{\Omega}$ and $\mathbf{\Omega}^\star$.
\\[2mm]
We use the average $\bar{\mathcal{D}}(t)$ in Eq. (\ref{Daver_general}) and its maximum value $\bar{\mathcal{D}}_{\mathrm{max}}$ in Eq. (\ref{Dmax_aver_def}) for the comparison of the synchronization processes generated by $\mathbf{\Omega}$ and $\mathbf{\Omega}^\star$. In this case, it is important to notice that the evaluation of $\bar{\mathcal{D}}_{\mathrm{max}}$ requires calculating each $\mathcal{D}_s(t)$ for the nodes $s=1,2,\ldots,N$, having to solve numerically the Kuramoto model with the initial condition $\theta_s(0)=\theta^\star_s(0)=\theta_0$, $\theta_m(0)=\theta^\star_m(0)=0$ for $m=1,\ldots,N$ and $m\neq s$. In our analysis, we choose $\theta_0=\pi/2$.  The numerical results for networks  with different $N$ are shown in Fig. \ref{Fig_4}. In Figs. \ref{Fig_4}(a) to  \ref{Fig_4}(c) we present  $\bar{\mathcal{D}}(t)$ as a function of $t$, the results are shown as different curves generated for $2\leq \ell\leq\lfloor N/2\rfloor$ codified in the colorbar as $\ell/N$. The numerical values of $\bar{\mathcal{D}}(t)$ show that $\bar{\mathcal{D}}(t)\approx 0 $ for $t$ small, then $\bar{\mathcal{D}}(t)$ gradually increases. For  $\ell\gg 2$ we observe a peak that rises with $\ell$. The results are in good agreement with the fact that introducing a chord with small $\ell$, the averages of $\mathcal{D}_s(t)$ defined in Eq. (\ref{def_D_s}) over all the nodes $s$ are small since the chord only produces little variations affecting the global dynamics. Furthermore, $\ell$ large creates greater connectivity that substantially changes the synchronization with respect to the original ring. The results show that, for $\ell$ large, the time $t$ where is produced the maximum of $\bar{\mathcal{D}}(t)$  increases with the size of the network. However, in the limit $t\to \infty$, $\bar{\mathcal{D}}_\infty\to 0$ due to the fact that $\mathbf{\Omega}$ and $\mathbf{\Omega}^\star$ are symmetric. 
\\[2mm]
In Fig. \ref{Fig_4}(d) we present the values of the maximum dissimilarity $\bar{\mathcal{D}}_{\mathrm{max}}$ in Eq. (\ref{Dmax_aver_def}) in terms of the value $\ell/N$. The results allow us to compare the dynamics on the ring and the effect of the chord. For the different sizes of the networks, we see $\bar{\mathcal{D}}_{\mathrm{max}}$ increases monotonically with $\ell$. The findings for the curves for $\bar{\mathcal{D}}_{\mathrm{max}}$ are similar for the sizes $N\geq 50$, and only small deviations are observed for the networks with $N=10$ and $N=20$. This result suggests that the effect of a chord on large rings produces differences in the transitory states with modifications in the synchronization depending only on $\ell/N$.
\subsection{Synchronization on graphs with $N=4$}
\begin{figure*}[t]
	\centering
	\includegraphics*[width=0.9\textwidth]{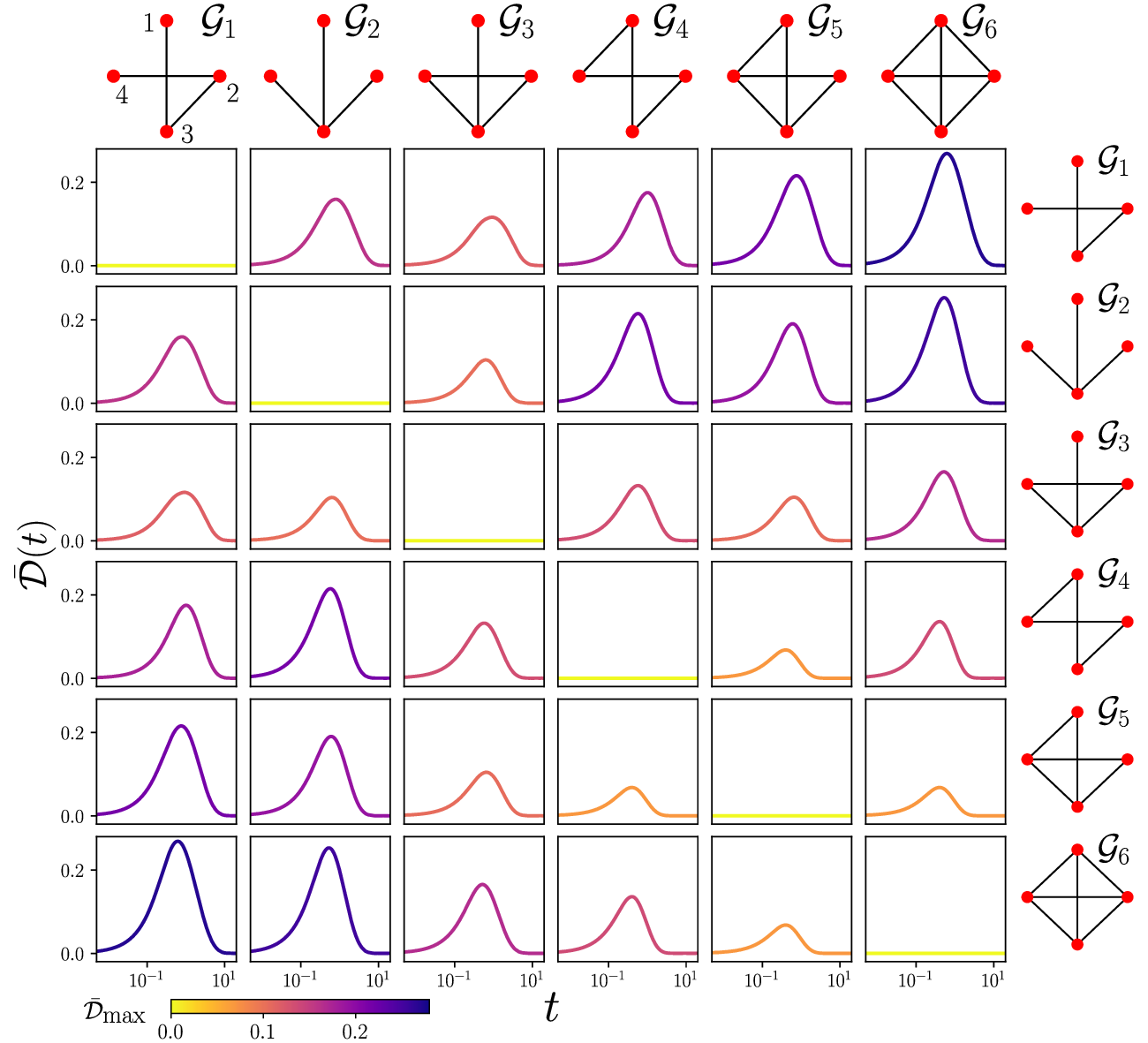}
	\vspace{-2mm}
	\caption{\label{Fig_5} Comparison of synchronization on connected nonisomorphic graphs with $N=4$. We compare the dynamics defined by $\mathbf{\Omega}$ and $\mathbf{\Omega}^\star$ as the respective adjacency matrices for the graphs denoted as  $\mathcal{G}_1,\,\mathcal{G}_2,\ldots,\mathcal{G}_6$ and obtained from Ref. \cite{ConnectedGraphs}. We present $\bar{\mathcal{D}}(t)$ as a function of $t$, the results are generated with the numerical evaluation of Eq.  (\ref{Daver_general}). $\bar{\mathcal{D}}_{\mathrm{max}}$ in Eq. (\ref{Dmax_aver_def}) is used to color each curve (see details in the main text).}
\end{figure*}
In the previous sections, we explored synchronization using a reference network and the effect of  modifications to this structure. However, the approach introduced is general and can be applied to detect differences in synchronization processes between any pair of connected networks with the same size. To illustrate this, we compare the dynamics between all nonisomorphic connected networks with $N=4$ nodes. The graphs are denoted by $\mathcal{G}_1,\, \mathcal{G}_2,\ldots, \mathcal{G}_6$ and shown in Fig. \ref{Fig_5}, this set was obtained from Ref. \cite{ConnectedGraphs}. These particular networks include several topologies of interest: $\mathcal{G}_1$ is a linear graph, $\mathcal{G}_2$ is a tree known as a star graph, $\mathcal{G}_3$ is formed by a clique with three nodes (triangle) and an additional node connected by one edge, $\mathcal{G}_4$ is a ring, $\mathcal{G}_5$ is a graph formed by two triangles sharing an edge and $\mathcal{G}_6$ is a fully connected graph.
\\[2mm]
In Fig. \ref{Fig_5}, we compare the synchronization in all the graphs $\mathcal{G}_1,\, \mathcal{G}_2,\ldots, \mathcal{G}_6$. We consider pairs of graphs $\mathcal{G}_\nu$, $\mathcal{G}_\rho$ with $\nu,\rho=1,2,\ldots, 6$, where $\nu$ denotes the rows and $\rho$ the columns in the plot panels. At the top and right borders, we show the graphs used; also, as a reference, in  $\mathcal{G}_1$ at the top are included the node labels $1,\,2,\,3,\,4$, this layout is maintained in all the graphs presented in the figure. We calculate $\bar{\mathcal{D}}(t)$ in Eq. (\ref{Daver_general})  using the adjacency matrix (with elements $1$ if two nodes are connected and $0$ otherwise) of $\mathcal{G}_\nu$ to define $\mathbf{\Omega}$ and the adjacency matrix of $\mathcal{G}_\rho$ to define $\mathbf{\Omega}^\star$. The numerical solution of Eq. (\ref{kuramoto_omega}) for both coupling matrices allows to calculate $\mathcal{D}_s(t)$ with the initial condition $\theta_s(0)=\theta_s^\star(0)=\theta_0=\pi/2$ and $\theta_\ell(0)=\theta_\ell^\star(0)=0$ for all $\ell=1,\, 2,\, 3,\, 4$ and $\ell\neq s$. All this information is combined in Eq. (\ref{Daver_general}) for all the nodes $s$ to obtain $\bar{\mathcal{D}}(t)$ as a function of $t$. We repeat this procedure for all pairs $\nu,\rho$ to generate the 36 curves depicted in the panels in Fig. \ref{Fig_5}, all these curves are colored with the respective $\bar{\mathcal{D}}_{\mathrm{max}}$ in Eq. (\ref{Dmax_aver_def})  codified in the colorbar. In all the cases studied, it is observed that $\bar{\mathcal{D}}_{\mathrm{max}}$ is produced by phase separations in transitory states.
\\[2mm]
In the first row in Fig. \ref{Fig_5}, we show the results for the comparison of the synchronization using as reference the dynamics on $\mathcal{G}_1$. The numerical values for $\bar{\mathcal{D}}_{\mathrm{max}}$ reveal that the most similar process occurs in $\mathcal{G}_3$, the respective graphs differ only in the edge (3,4). Furthermore, in the second row, when the graph  $\mathcal{G}_2$ is used as reference the dynamics is also more affine with $\mathcal{G}_3$, both graphs differ in the additional edge (2,4) present in $\mathcal{G}_3$. In contrast, when we use as reference the dynamics on $\mathcal{G}_3$ (panels in the third row), the more similar dynamics are observed in $\mathcal{G}_2$ and $\mathcal{G}_5$, structures that differ with $\mathcal{G}_3$ deleting the link (2,4) or adding (1,4). The analyses displayed in the fourth row using as reference the ring $\mathcal{G}_4$ show that the most similar process is produced in $\mathcal{G}_5$ whereas the maximum value $\bar{\mathcal{D}}_{\mathrm{max}}$ is found when the ring $\mathcal{G}_4$ is compared with the star graph $\mathcal{G}_2$. However, the comparison of $\mathcal{G}_5$ (panels in the fifth row) evidence similarities with $\mathcal{G}_4$ and $\mathcal{G}_6$ obtained removing from $\mathcal{G}_5$ the edge (3,4) or adding the link (1,2).
\\[2mm]
Finally, when we compare the synchronization using the fully connected graph $\mathcal{G}_6$ as reference (panels in the sixth row in Fig. \ref{Fig_5}) with the rest of the graphs, we see that the values $\bar{\mathcal{D}}_{\mathrm{max}}$ are sorted in decreasing order from $\mathcal{G}_1$ to $\mathcal{G}_5$ showing that the greatest differences appear between the dynamics on $\mathcal{G}_6$ and the linear graph $\mathcal{G}_1$. The order observed evidences that the approach developed detects and quantifies differences in the synchronization on networks. For example, if we compare only the coupling matrices (which in this case are the adjacency matrices), it is impossible to establish which of the processes in $\mathcal{G}_1$ or in $\mathcal{G}_2$ is more similar to the dynamics in $\mathcal{G}_6$ since both $\mathcal{G}_1$ and $\mathcal{G}_2$ have the same number of links and $\sum_{i,j}(\Omega^\star_{ij}-\Omega_{ij})=6$ in both cases; however, the results show that between $\mathcal{G}_1$ and $\mathcal{G}_2$, the synchronization in  $\mathcal{G}_2$ is more similar to  $\mathcal{G}_6$. Something analogous happens with  $\mathcal{G}_3$ and  $\mathcal{G}_4$ when compared with  $\mathcal{G}_6$, if we evaluate the differences between matrix elements we have  $\sum_{i,j}(\Omega^\star_{ij}-\Omega_{ij})=4$ since the mentioned graphs differ in two links with $\mathcal{G}_6$; but, in this case a lower $\bar{\mathcal{D}}_{\mathrm{max}}$ is found when comparing the ring $\mathcal{G}_4$ with the complete graph $\mathcal{G}_6$. Finally, it is also observed that the synchronization dynamics are more similar between $\mathcal{G}_5$ and $\mathcal{G}_6$ as the graphs differ in only one edge.
\subsection{Comparing the Kuramoto model with its linear approximation}
\begin{figure*}[t]
	\centering
	\includegraphics*[width=0.95\textwidth]{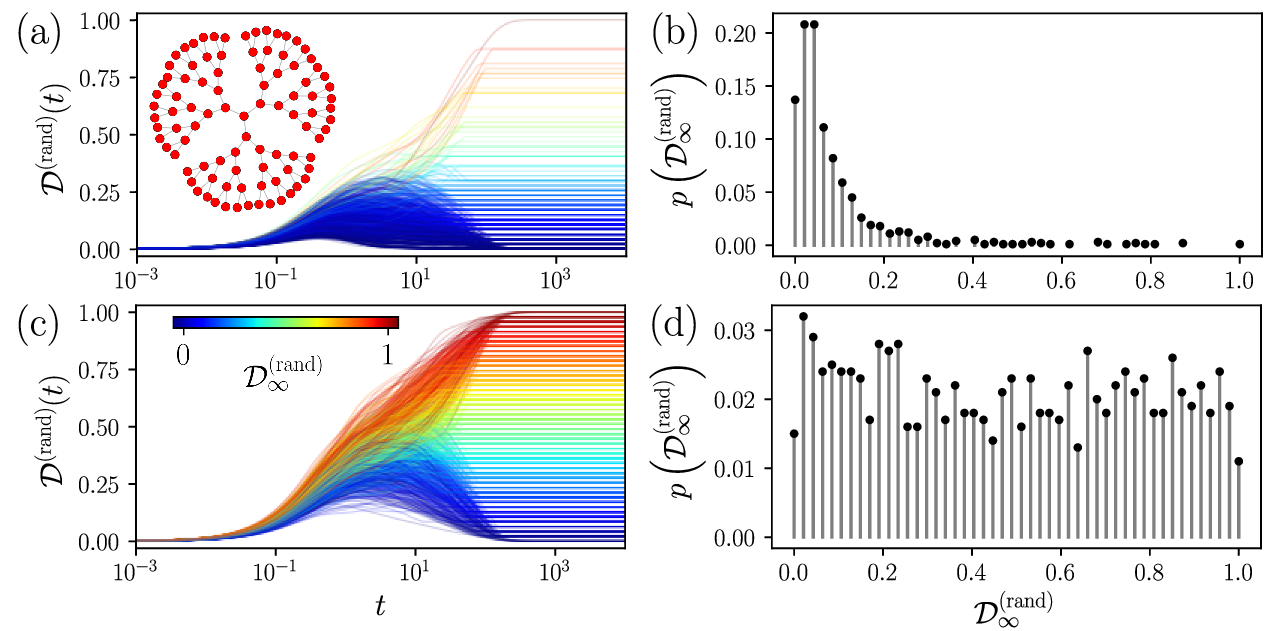}
	\vspace{-5mm}
	\caption{\label{Fig_6} Differences between the dynamics with the Kuramoto model and the linear approximation on a Cayley tree with $N=94$ nodes using initial random phases. (a) $\mathcal{D}^{(\mathrm{rand})}(t)$ as a function of $t$, the results are obtained numerically considering random phases at $t=0$ with average $\theta_0=3\pi/4$, the different curves depict 1000 realizations. The Cayley tree analyzed is presented as inset. (b) Statistical analysis of the values $\mathcal{D}^{(\mathrm{rand})}(t)$ for $t$ large, denoted as $\mathcal{D}^{\mathrm{(\mathrm{rand})}}_\infty$, the relative frequency $p\left(\mathcal{D}^{\mathrm{(\mathrm{rand})}}_\infty\right)$ is obtained from the 1000 values in (a) for $t=10^4$. Panels (c) and (d) show the same analysis for random initial phases with average $\theta_0=\pi$. In (a) and (d) the curves are presented with different colors according to $\mathcal{D}^{\mathrm{(\mathrm{rand})}}_\infty$ codified in the colorbar in (c). }
\end{figure*}
Our previous examples emphasized in the comparison of the dynamics of coupled identical oscillators evolving with the Kuramoto model considering  two coupling matrices $\mathbf{\Omega}$ and $\mathbf{\Omega}^\star$. However, the approach explored in Sec. \ref{Sec_Definitions} can be generalized to analyze other situations. In the following, we compare two dynamics in the same network; now, the reference process is the system of coupled oscillators evolving with the Kuramoto model defined in Eq. (\ref{kuramoto_omega}) and the second dynamics is determined by the linear approximation in Eq. (\ref{dtheta_linear}) maintaining the same coupling matrix $\mathbf{\Omega}=\mathbf{A}$ defined in terms of the adjacency matrix. In addition, to see the response of the processes to the initial conditions, the phases of the reference process $\{\theta_m(t)\}_{m=1}^N$ as well as the second process with phases $\{\theta_m^\star(t)\}_{m=1}^N$ have the same initial condition $\theta_m(0)=\theta_m^\star(0)=\theta_0^{(m)}$ for $m=1,2,\ldots,N$, where each $\theta_0^{(m)}$ is a random variable uniformly distributed in such a way that the average of the initial phases is $\theta_0$, 
i.e., the initial condition satisfies
\begin{equation}\label{theta_0_rand}
\theta_0=\frac{1}{N}\sum_{m=1}^N\theta_0^{(m)}.
\end{equation}
To obtain the values  $\theta_0^{(m)}$, we generate $N$ random values uniformly distributed in the interval $(0,1]$, then we divide each value between the average, and each result is multiplied by $\theta_0$ to fulfill Eq. (\ref{theta_0_rand}).
\\[2mm]
Once generated the initial condition for the oscillators, the Kuramoto model in Eq. (\ref{kuramoto_omega}) and the linear approximation in Eq. (\ref{dtheta_linear}) are solved numerically. The values allow calculating the distance between phases defined as
\begin{equation}\label{D_rand_t}
\mathcal{D}^{(\mathrm{rand})}(t)\equiv\frac{1}{N \pi}\sum_{\ell=1}^N d\left[\theta_\ell(t)-\theta^\star_\ell(t)\right].
\end{equation}
\begin{figure*}[t]
	\centering
	\includegraphics*[width=0.95\textwidth]{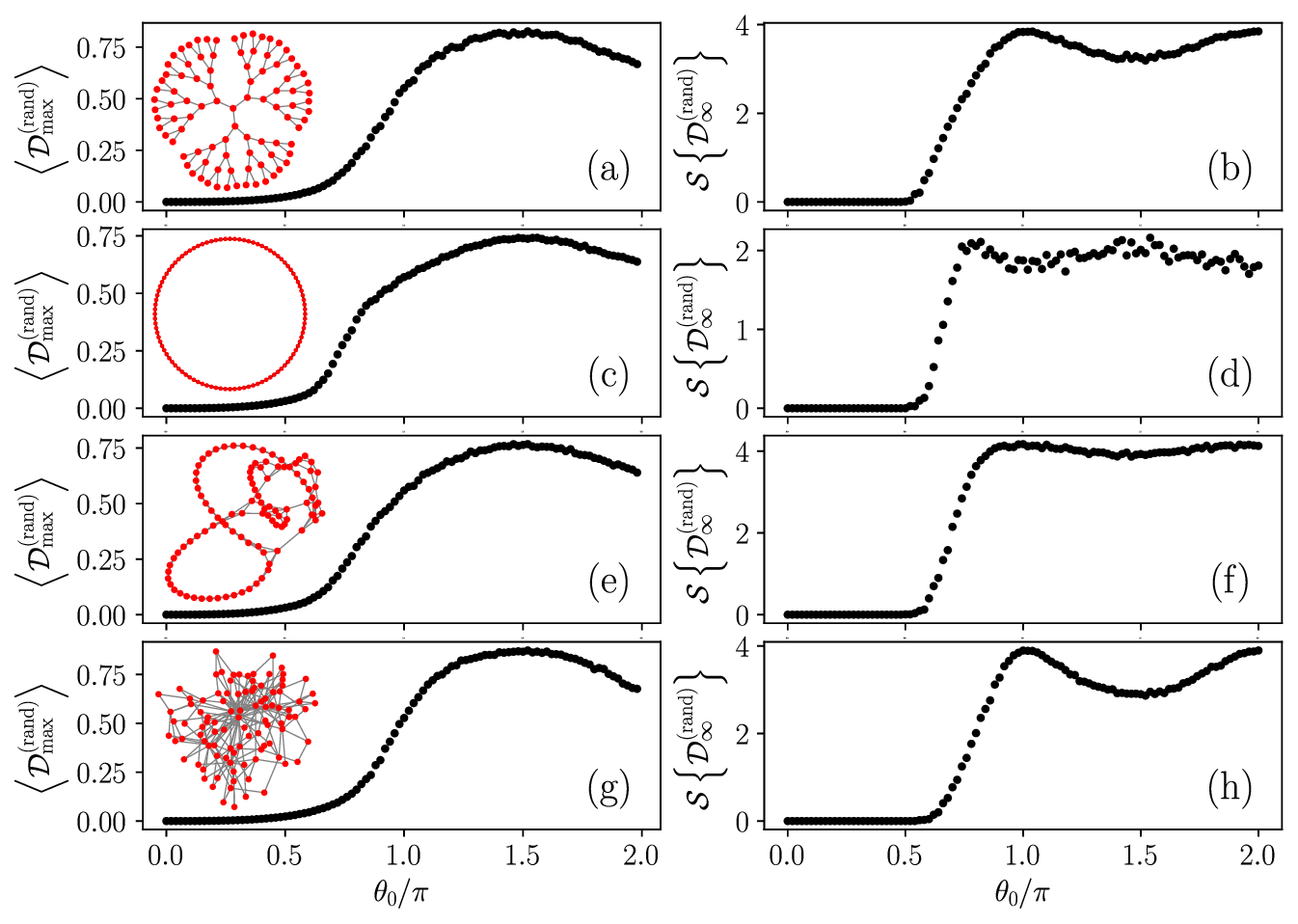}
	\vspace{-5mm}
	\caption{\label{Fig_7} Differences between the Kuramoto model and the linear approximation in networks. Numerical values of $\langle\mathcal{D}^{(\mathrm{rand})}_\mathrm{max}\rangle$ (left panels) and $\mathcal{S}\left\{\mathcal{D}^{(\mathrm{rand})}_\infty\right\}$ (right panels) as a function of the average phases of the initial conditions $\theta_0$ for: (a), (b) the Cayley tree studied in Fig. \ref{Fig_6}, (c), (d)  an interacting cycle with degree $k=4$, (e), (f)  a Watts-Strogatz random network with rewiring probability $p=0.02$, (g), (h) a Barab\'asi-Albert  random network, where each newly introduced node connects to $m=2$ previous nodes. The networks are depicted as inset in the left panels. The results are obtained using 1000 realizations for each value of $\theta_0$.}
\end{figure*}
Using this definition $\mathcal{D}^{(\mathrm{rand})}(t)$ depends on the initial phases and evolves according to the two processes to be compared. In relation with the definition in Eq. (\ref{def_D_s}), $\mathcal{D}^{(\mathrm{rand})}(t)$ includes the additional factor $1/N$ since in Eq. (\ref{D_rand_t}) all the initial phases are different from zero whereas in $\mathcal{D}_s(t)$ in Eq. (\ref{def_D_s}) only the phase of node $s$ is non-null.
\\[2mm]
In Fig. \ref{Fig_6} we analyze $\mathcal{D}^{(\mathrm{rand})}(t)$ for a Cayley tree with $N=94$ nodes generated with the coordination number $z=3$ and $n=5$ shells, the network is presented as inset in Fig. \ref{Fig_6}(a). In  Fig. \ref{Fig_6}(a) we show $\mathcal{D}^{(\mathrm{rand})}(t)$ as a function of $t$ using 1000 realizations of the initial conditions with average $\theta_0=3\pi/4$ in Eq. (\ref{theta_0_rand}). The results show different behaviors for  $\mathcal{D}^{(\mathrm{rand})}(t)$; in particular, for $t$ large,
$\mathcal{D}^{\mathrm{(\mathrm{rand})}}_\infty\equiv \lim_{t\to \infty}\mathcal{D}^{(\mathrm{rand})}(t)$ presents values distributed between 0 and 1 revealing that the two dynamical processes (Kuramoto model and linear approximation)  produce different stationary phases in cases where $\mathcal{D}^{\mathrm{(\mathrm{rand})}}_\infty>0$. We colored the curves with $\mathcal{D}^{\mathrm{(\mathrm{rand})}}_\infty$. In Fig. \ref{Fig_6}(b), we analyze statistically the values $\mathcal{D}^{\mathrm{(\mathrm{rand})}}_\infty$ for the 1000 realizations, the probabilities  $p\left(\mathcal{D}^{\mathrm{(\mathrm{rand})}}_\infty\right)$  in the bars show the relative frequency of $\mathcal{D}^{\mathrm{(\mathrm{rand})}}_\infty$ obtained from $\mathcal{D}^{(\mathrm{rand})}(t)$ in Fig. \ref{Fig_6}(a) for $t=10^4$ (the frequency counts are made using the values rounded with three decimals). This representation help us to see the differences in the configurations when the oscillators reach stationary states. In Figs. \ref{Fig_6}(c) and \ref{Fig_6}(d), we repeat the analysis of $\mathcal{D}^{(\mathrm{rand})}(t)$ for initial conditions with the average value $\theta_0=\pi$. We see how the increase of $\theta_0$ generates more diversity in the results for the curves $\mathcal{D}^{(\mathrm{rand})}(t)$ with respect to the observed in Figs.\ref{Fig_6}(a) and \ref{Fig_6}(b); in particular, for the final values $\mathcal{D}^{\mathrm{(\mathrm{rand})}}_\infty$.
\\[2mm]
To compare the diversity of the values $\mathcal{D}^{\mathrm{(\mathrm{rand})}}_\infty$ found in  Fig. \ref{Fig_6}, we use the Shannon's entropy $\mathcal{S}$ that for the probabilities $\{p_m\}_{m=1}^M$ of $M$ discrete values is defined by \cite{Shannon_1948}
\begin{equation}\label{Shannon_entropy}
\mathcal{S}\equiv -\sum_{m=1}^M p_m\ln p_m
\end{equation}
and quantifies the diversity of the dataset described with the probabilities $p_m$ fulfilling $\sum_{m=1}^Mp_m=1$ and $0<p_m\leq 1$. The analysis of the probabilities found in Fig. \ref{Fig_6}(b) produces the value $\mathcal{S}=2.38$ whereas for Fig. \ref{Fig_6}(d) $\mathcal{S}= 3.85$, giving numerical evidence of the changes of $\mathcal{D}^{\mathrm{(\mathrm{rand})}}_\infty$ in the Monte Carlo simulations when the average phases in the initial conditions changes.
\\[2mm]
The results in Fig. \ref{Fig_6} evidence the diversities in the evolution of phases in $\mathcal{D}^{(\mathrm{rand})}(t)$ and how they are modified with the initial conditions of the oscillators. The findings also suggest that the ensemble average of the maximum of $\mathcal{D}^{(\mathrm{rand})}(t)$ denoted as $\left\langle \mathcal{D}^{\mathrm{(\mathrm{rand})}}_{\mathrm{max}}\right\rangle$ as a function of the average phases of oscillators $\theta_0$ could help to understand the effect of this quantity in the differences of the Kuramoto model and the linear approximation. This result is complemented by the entropy defined in Eq. (\ref{Shannon_entropy}) associated with the probabilities $p\left(\mathcal{D}^{\mathrm{(\mathrm{rand})}}_\infty\right)$ of the values $\mathcal{D}^{\mathrm{(\mathrm{rand})}}_\infty$ and denoted as $\mathcal{S}\left\{\mathcal{D}^{\mathrm{(\mathrm{rand})}}_\infty\right\}$. In Fig. \ref{Fig_7} we show the numerical values of $\left\langle \mathcal{D}^{\mathrm{(\mathrm{rand})}}_{\mathrm{max}}\right\rangle$ (presented in the left panels) and $\mathcal{S}\left\{\mathcal{D}^{\mathrm{(\mathrm{rand})}}_\infty\right\}$ (right panels) as a function of $\theta_0$ for four networks. The results are obtained from 1000 realizations implementing the same approach illustrated in Fig. \ref{Fig_6} considering the coupling matrix $\mathbf{\Omega}$ defined by the adjacency matrix $\mathbf{A}$ of each network for two deterministic networks [Figs. \ref{Fig_7}(a)-\ref{Fig_7}(d)] and two random networks [Figs. \ref{Fig_7}(e)-\ref{Fig_7}(h)].
\\[2mm]
In Figs. \ref{Fig_7}(a) and \ref{Fig_7}(b), we analyze the synchronization in the Cayley tree with $N=94$ explored in Fig. \ref{Fig_6}. Now we study the effect of $\theta_0$ on $\left\langle \mathcal{D}^{\mathrm{(\mathrm{rand})}}_{\mathrm{max}}\right\rangle$ [Fig. \ref{Fig_7}(a)] and the entropy $\mathcal{S}\left\{\mathcal{D}^{\mathrm{(\mathrm{rand})}}_\infty\right\}$ [Fig. \ref{Fig_7}(b)], the results for the last quantity agree with the values obtained for the cases with $\theta_0=3\pi/4$ and $\theta_0=\pi$ deduced from the probabilities in Figs. \ref{Fig_6}(b) and \ref{Fig_7}(d). In Figs. \ref{Fig_7}(c) and \ref{Fig_7}(d), the coupling is determined by an interacting cycle with $N=100$ defined by a ring and additional edges connecting nodes at distance 2 in the original ring (the adjacency matrix $\mathbf{A}$ is a circulant matrix with the non-null entries $c_1=c_{N-1}=1$ and $c_2=c_{N-2}=1$), the degree of each node is $k=4$. In Figs. \ref{Fig_7}(e) and \ref{Fig_7}(f) is analyzed the Watts–Strogatz network created from the regular structure in Fig.  \ref{Fig_7}(c) and rewiring it randomly with probability $p=0.02$, the reorganization of the edges reduces the average shortest path length between nodes inducing the small-world property \cite{WattsStrogatz1998}. In Figs. \ref{Fig_7}(g) and \ref{Fig_7}(h) are studied the coupled dynamics on a scale-free Barab\'asi-Albert random network generated with the preferential attachment rule where each newly introduced node connects to $m=2$ previous nodes \cite{BarabasiAlbert1999}. In this network the degrees have different values with the existence of a few hubs and a high fraction of nodes with few neighbors.
\\[2mm]
The results in Fig. \ref{Fig_7} show the dissimilarities measured with the ensemble averages of $\mathcal{D}^{\mathrm{(\mathrm{rand})}}_\mathrm{max}$ obtained from $\mathcal{D}^{(\mathrm{rand})}(t)$. For all the networks analyzed, it is found that $\langle \mathcal{D}^{\mathrm{(\mathrm{rand})}}_\mathrm{max} \rangle$ is small for average phases of the initial conditions satisfying $\theta_0\leq \pi/2$, then $\langle \mathcal{D}^{\mathrm{(\mathrm{rand})}}_\mathrm{max} \rangle$ increases gradually in the interval $\pi/2<\theta_0<3\pi/2$ reaching a maximum around $\theta_0\approx 3\pi/2$. It is also observed that after this maximum, $\langle \mathcal{D}^{\mathrm{(\mathrm{rand})}}_\mathrm{max} \rangle$ decreases because, for values close to $\theta_0=2\pi$, groups of oscillators have initial phases for which the linear approximation is valid, thus reducing $\langle \mathcal{D}^{\mathrm{(\mathrm{rand})}}_\mathrm{max} \rangle$.
\\[2mm]
In addition, although qualitatively the curves for $\langle \mathcal{D}^{(\mathrm{rand})}_\mathrm{max} \rangle$  as a function of $\theta_0$ are similar when compared in detail they present some differences associated with the topologies of the networks. For example, the largest changes in $\langle \mathcal{D}^{(\mathrm{rand})}_\mathrm{max} \rangle$ are observed for the Cayley tree and the Barab\'asi-Albert network.
\\[2mm]
Furthermore, the entropy obtained from the statistical analysis of $\mathcal{D}^{(\mathrm{rand})}_\infty$ when each system has reached its steady state shows an abrupt change around $\theta_0 \approx \pi/2$ with cases in which $\mathcal{S}\left\{\mathcal{D}^{(\mathrm{rand})}_\infty\right\}=0$ for $\theta_0 \leq \pi/2$ showing that the phases for $t\to \infty$ coincide in the Kuramoto model and the linear approximation but that in the interval $\pi/2<\theta_0\leq \pi$ exhibit a rapid change (somewhat comparable to a phase transition between similar and dissimilar dynamics). In the interacting cycle [Fig. \ref{Fig_7}(d)] and the Watts-Strogatz network [Fig. \ref{Fig_7}(f)], it is observed that $\mathcal{S}\left\{\mathcal{D}^{(\mathrm{rand})}_\infty\right\}$ remains close to a constant for average angles $\theta_0> \pi$. While it is very striking that in the Cayley tree [Fig.   \ref{Fig_7}(b)] and the Barab\'asi-Albert network [Fig. \ref{Fig_7}(h)] the entropy values vary in the interval $\theta_0>\pi$ presenting a local minimum at $\theta_0\approx 3\pi/2$.
\\[2mm]
In general, all the results obtained in this section considering the effect of the initial conditions in several topologies open new questions such as what mechanisms generate the phase separation in the Kuramoto model and its linear approximation and show us the usefulness and potential that a measure of dissimilarity between two synchronization processes can have.
\\[3mm]
\section{Conclusions}
\label{Sec_Conclusions}
In this study, we introduced a formalism to compare synchronization processes occurring in systems of $N$ oscillators coupled by networks. The explored metric considers the differences in the oscillator phases between two processes in a space described by an $N$-dimensional hypertorus, quantifies the effect of the initial conditions using a dynamical process as a reference, and compares it with a second process.
\\[2mm]
We tested the methods introduced in the analysis of different systems of identical coupled oscillators that evolve with the Kuramoto model as well as its linear approximation. In particular, we study the non-symmetric coupling of two oscillators, processes described by circulant matrices, the effect of adding one edge in a ring, and the comparison of synchronization in all connected graphs with four nodes. The results show that $\mathcal{\bar{D}}(t)$ defined in Eq. (\ref{Daver_general}) and its maximum $\bar{\mathcal{D}}_{\mathrm{max}}$ characterize the differences between the two synchronization processes. Finally, the approach was adapted to evaluate the differences between the Kuramoto model and its linear approximation, considering the same network and random initial phases. These dynamics were explored in two deterministic and two random networks.
\\[2mm]
The methods developed are general and can be applied to a diverse variety of coupled systems, for example, considering systems with different oscillation frequencies, to see the effect produced by synchronization models with nonlinear functions more general than the Kuramoto model and to characterize the consequences of noise with the incorporation of stochastic terms in coupled differential equations, among many other cases. The entire approach provides insights into the information to be contemplated when comparing dynamic processes occurring in complex systems and paves the way for new metrics combining the structure of a system as well as the processes occurring on it.
\onecolumngrid

\end{document}